\documentclass[twocolumn,showpacs,amsmath,amssymb,showkeys,noshowpacs,prc]{revtex4}
\usepackage{mathrsfs}
\usepackage{graphicx,color}
\usepackage{multirow}
\usepackage{dcolumn}
\usepackage{bm}
\usepackage{CJK}

\begin{document}
\title{High precision nuclear mass predictions towards a hundred kilo-electron-volt accuracy}

\author{Zhongming Niu$^{a}$}
\author{Haozhao Liang$^{b}$}
\author{Baohua Sun$^{c}$}
\author{Yifei Niu$^{d}$}
\author{Jianyou Guo$^{a}$}
\author{Jie Meng$^{c,e,f}$}
\affiliation{$^a$School of Physics and Materials Science, Anhui University, Hefei 230601, China}
\affiliation{$^b$RIKEN Nishina Center, Wako 351-0198, Japan}
\affiliation{$^c$School of Physics and Nuclear Energy Engineering, Beihang University,
             Beijing 100191, China}
\affiliation{$^d$ELI-NP, ``Horia Hulubei" National Institute for Physics and Nuclear Engineering, 30 Reactorului Street, RO-077125 Bucharest-Magurele, Romania}
\affiliation{$^e$State Key Laboratory of Nuclear Physics and Technology, School of Physics, Peking
             University, Beijing 100871, China}
\affiliation{$^f$Department of Physics, University of Stellenbosch,
             Stellenbosch 7602, South Africa}

\date{\today}

\begin{abstract}
Mass is a fundamental property and an important fingerprint of atomic nucleus. It provides an extremely useful test ground for nuclear models and is crucial to understand energy generation in stars as well as the heavy elements synthesized in stellar explosions. Nuclear physicists have been attempting at developing a precise, reliable, and predictive nuclear model that is suitable for the whole nuclear chart, while this still remains a great challenge even in recent days. Here we employ the Fourier spectral analysis to examine the deviations of nuclear mass predictions to the experimental data and to present a novel way for accurate nuclear mass predictions. In this analysis, we map the mass deviations from the space of nucleon number to its conjugate space of frequency, and are able to pin down the main contributions to the model deficiencies. By using the radial basis function approach we can further isolate and quantify the sources. Taking a pedagogical mass model as an example, we examine explicitly the correlation between nuclear effective interactions and the distributions of mass deviations in the frequency domain. The method presented in this work, therefore, opens up a new way for improving the nuclear mass predictions towards a hundred kilo-electron-volt accuracy, which is argued to be the chaos-related limit for the nuclear mass predictions.
\end{abstract}



\maketitle

\section{Introduction}
Nucleus is a very small but dense object in an atom. It contains almost all mass of an atom, and hence becomes the major origin of masses in our world. Known as the famous Einstein mass-energy relation $E=mc^2$, the tiny “mass defect” in atomic nuclei gives the energies that fuel the stars, including our sun, and further provides the energies for the lives on the earth~\cite{Bethe1936PR}. Moreover, nuclear mass is a key nuclear physics input for nuclear astrophysics in understanding the origin of elements in our Universe~\cite{Burbidge1957RMP}, the composition of the most compact objects known-neutron stars~\cite{Wolf2013PRL}, the neutrino cooling of the neutron star crusts~\cite{Schatz2014Nature}, etc. Due to the lack of nuclear properties for most nuclei related to the rapid neutron capture process ($r$-process), the origin of heavy elements is still an unsolved physics question~\cite{Qian2007PRp, Arnould2007PRp}, which is one of the $11$ greatest unanswered questions of physics~\cite{Haseltin2002Discover}. High-precision determination of nuclear mass has always been extremely important for nuclear physicists~\cite{Lunney2003RMP, Bender2003RMP, Meng2006PPNP, Bao2017SCPMA}, since it plays an essential role in determining the limits of nuclear landscape~\cite{Erler2012Nature}, understanding the nuclear interaction and the behavior of neutron-rich matter~\cite{Wienholtz2013Nature, Yu2016SCPMA}, and studying the nuclear shell structure~\cite{Ramirez2012Science}.

With the construction and upgrade of radioactive ion beam facilities, great progress has be made in recent years in the measurements of nuclear masses, which have reached very high precision~\cite{Blaum2006PRp, Rainville2004Science, Block2010Nature, Ramirez2012Science} and have been available for very exotic nuclei~\cite{Franzke2008MSR, Sun2015FP, Wienholtz2013Nature, Zhang2012PRL, Tu2011PRL}. The nucleus is a finite quantum many-body system that is composed of two types of interacting fermions in which the underlying force is poorly understood. Thus the prediction of nuclear mass is a great and longstanding challenge for theoretical models. The accurate mass prediction is largely hampered by the absence of an exact theory of nuclear interaction and the difficulties inherent to quantum many-body calculations. Therefore, various models have been or are being developed to predict nuclear masses. Although the $ab~initio$ calculations can be used to predict the nuclear masses, they are only applicable to the light nuclei or those nuclei near magic numbers~\cite{Barrett2013PPNP, Pieper2001ARNPS, Liu2012PRC, Soma2013PRC, Morris2018PRL}. For the whole nuclear chart, a classical nuclear mass model --- liquid drop model --- was developed~\cite{Bethe1936RMP, Weizsacker1935ZP} soon after the confirmation of the constituents of nucleus, i.e., the discovery of neutron~\cite{Chadwick1932Nature}. Since then, comprehensive efforts have been devoted to this investigation. There are so far mainly three kinds of global nuclear mass models: the macroscopic [e.g., Bethe-Weizs\"{a}cker (BW) model~\cite{Bethe1936RMP, Weizsacker1935ZP, Kirson2008NPA}], macroscopic-microscopic [e.g., finite-range droplet model (FRDM)~\cite{Moller1995ADNDT} and Weizs\"{a}cker-Skyrme (WS) model~\cite{Wang2014PLB}], and microscopic mass models [e.g., Skyrme Hartree-Fock-Bogoliubov (HFB)~\cite{Goriely2013PRC} and relativistic mean-field (RMF) models~\cite{Geng2005PTP,Hua2012SCPMA}]. The macroscopic mass models well describe the bulk properties of nuclear mass, while they lack detailed information of nuclear shell structure and hence give relatively large root-mean-square (rms) deviations. By including the microscopic shell corrections, the macroscopic-microscopic mass models have achieved the best accuracy in nuclear mass descriptions, although this kind of hybrid models is generally inconsistent in their macroscopic and microscopic parts. The microscopic mass models are much involved, while they are usually believed to have a better ability of extrapolation.

About $80$ years' continuous efforts by several generations of theoreticians have resulted in a remarkable success in the development of nuclear models. The rms values, defined by the deviations between model predictions and experimental data~\cite{Wang2017CPC}, are reduced from about $3$ million electron volts (MeV) for the BW model~\cite{Kirson2008NPA} to about $300$ kilo electron volts (keV) for the WS4 model~\cite{Wang2014PLB}. Comparison of rms values in different models is shown in Fig. 1. The best accuracy of $300$-keV level achieved by the WS4 macroscopic-microscopic model~\cite{Wang2014PLB} is already a great challenge for the pure microscopic mass models, especially for those based on the covariant density functionals, which are usually believed to have a better ability of extrapolation. However, such an accuracy is still not good enough for the studies of exotic nuclear structure and astrophysical nucleosynthesis, which demand an accuracy better than $100$~keV~\cite{Lunney2003RMP}. From the estimate of statistical fluctuations of nuclear ground-state energies~\cite{Molinari2004PLB, Velazquez2005PLB}, the possible accuracy limit of theoretical calculations is expected to be tens of keV. Therefore, there are still a high demand and a large room for improving the existing nuclear mass models even for the best available one. Correspondingly, in this work, we would like to raise the following questions: what are the main limitations of present models? Microscopically, which parts of nuclear effective interactions are still heavily missing or need better treatment in the available theories? The answers to these questions are crucial for the breakthroughs in the coming studies in nuclear physics and astrophysics.

Based on thousands of measured nuclear masses~\cite{Wang2017CPC}, it has been aware that there are systematic large deviations from experimental data in various global mass models, e.g., around magic nuclei. Depending on specific model, these deviations often show quite different patterns in both magnitudes and distribution over the nuclear chart as a function of proton and neutron
numbers. In other words, it is difficult to find universal regulations that hold for different models for improving their accuracy. This motivates us to explore a novel view of the deviations between the calculated results and experimental data.

The Fourier spectral analysis provides a different view from the frequency domain, which has been widely used in many fields of engineering, such as electronics, telecommunication, and optics~\cite{Bracewell2000Book}. In particular, in the field of image processing, it is found that the frequency spectra of various pictures can show similar structures when they share common features. This makes the Fourier spectral analysis an effective way to find universal regulations of image processing, and thus plays an important role in image denoising, compression, and recognition. In fact, nuclear mass deviations between model predictions and experimental data can be treated as an image processing problem. We will first analyze the deviations by the Fourier spectral analysis in the conjugate space of frequency. As a step further, we will employ the radial basis function (RBF) approach to study the deviations coming from different frequency domains. The RBF approach is a powerful interpolation method, in which complicated nonlinear functions are described with linear combination of radial basis functions. It can well describe the smooth surface and has been successfully used in time series prediction, control of nonlinear systems, three-dimensional reconstruction in computer graphics~\cite{Buhmann2003Book}, even in nuclear mass predictions~\cite{Wang2011PRC, Niu2013PRCb, Zheng2014PRC}. Guided by the Fourier spectral analysis and the RBF approach, in this work, we will show a general and systematic strategy for improving the accuracy of nuclear mass predictions from several MeV to less than $100$ keV.

\section{Discrete Fourier transform}\label{Sec:2}
To investigate the deviations between the calculated results and experimental data in the conjugate frequency space, we would perform the two-dimensional discrete Fourier transform of the mass differences~\cite{Velazquez2004AIP}. The amplitude of discrete Fourier transform is defined as
\begin{eqnarray}
F_{kl} = \frac{1}{Z_{\rm{m}}N_{\rm{m}}}
         \sum_{Z=8}^{Z_{\rm{m}}}
         \sum_{N=8}^{N_{\rm{m}}}
         &&(M_{\rm{exp}}^{Z,N}-M_{\rm{th}}^{Z,N})\nonumber\\
         &&e^{-i2\pi\left[\frac{(k-1)(Z-1)}{Z_{\rm{m}}}
                       +\frac{(l-1)(N-1)}{N_{\rm{m}}}\right]},~~
\end{eqnarray}
where $i$ is the imaginary unit, $M_{\rm{exp}}^{Z,N}$ and $M_{\rm{th}}^{Z,N}$ are the experimental and theoretical nuclear masses, $Z_{\rm{m}}$ and $N_{\rm{m}}$ denote the maximum proton number and neutron number for the whole sample of nuclei. In the coordinate frame of proton frequency $\omega_\pi\equiv k/Z_{\rm{m}}$ and neutron frequency $\omega_\nu\equiv l/N_{\rm{m}}$: (i) the areas near the four corners correspond to the low $\omega_\pi$ and $\omega_\nu$ frequencies respectively; (ii) the areas near the centers of the four edges correspond to the high $\omega_\pi$ or $\omega_\nu$ frequencies respectively, and the area near the center corresponds to the high $\omega_\pi$ and $\omega_\nu$ frequencies; (iii) the other areas correspond to the $\omega_\pi$ and $\omega_\nu$ frequencies between low and high, we refer to the medium frequencies hereafter.

\section{Results and discussion}\label{Sec:3}
\begin{figure*}[!ht]
\includegraphics[width=16.6cm]{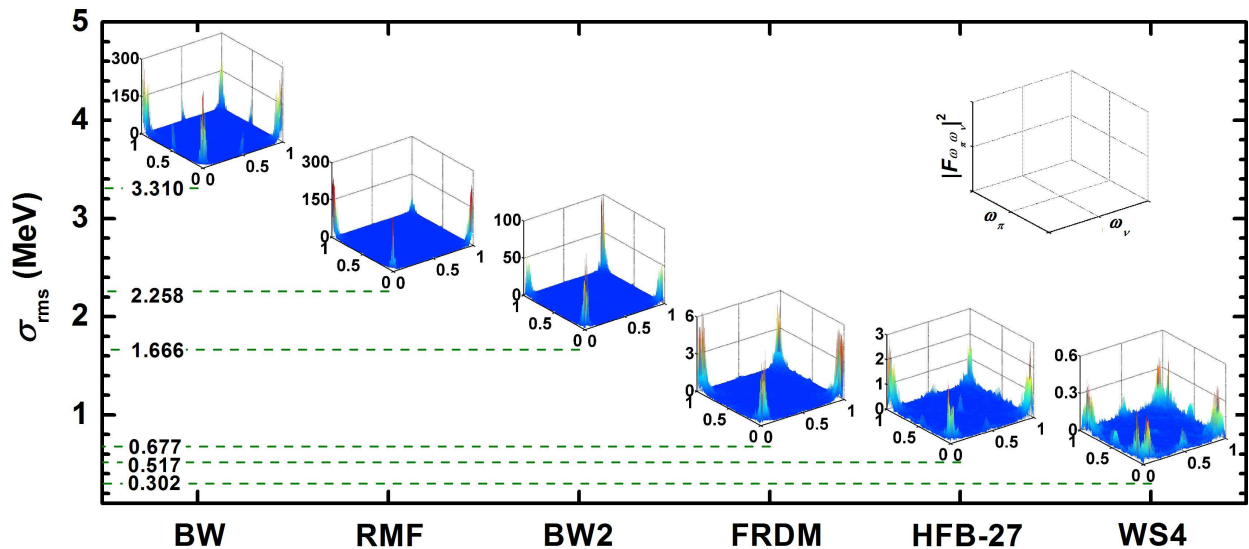}
\caption{(Color online) Squared amplitudes $|F_{\omega_\pi \omega_\nu}|^2$ of the discrete Fourier transforms of the mass differences between the experimental data and the predictions of various models, plotted against proton frequency $\omega_\pi$ and neutron frequency $\omega_\nu$. The rms deviations of these mass models are marked on the vertical axis. Here we only take into account the nuclei with proton number $Z\geqslant 8$ and neutron number $N\geqslant 8$ listed in the Atomic Mass Evaluation (AME2016).}
\label{fig:fig1}
\end{figure*}

Fig. 1 shows the squared amplitudes $|F_{\omega_\pi \omega_\nu}|^2$ of the two-dimensional discrete Fourier transforms of the mass differences for various mass models. The corresponding rms deviations are marked on the vertical axis. It is striking that the dominant peaks of $|F_{\omega_\pi \omega_\nu}|^2$ in all models identically locate at the low frequencies, although their distribution of mass deviations over the nuclear chart as a function of proton and neutron numbers are very different (see e.g., the patterns of BW2 and WS4 models in Fig. 2). Note the magnitude of $|F_{\omega_\pi \omega_\nu}|^2$ gradually decreases as the decrease of rms deviation, and meanwhile the high-frequency components gradually become visible. Eventually, for the WS4 model, the $|F_{\omega_\pi \omega_\nu}|^2$ at high frequencies are comparable to those at low frequencies. This implies that the physics details reflected by the peaks at high frequencies are revealing when the model accuracy approaches to better than $300$~keV. It is also noticed that a remarkable component is observed at the high-frequency domain for the BW model. This is due to the lack of pairing interaction in the BW model and the important insight will be clarified in the following.

\begin{figure*}[!ht]
\includegraphics[width=16.6cm]{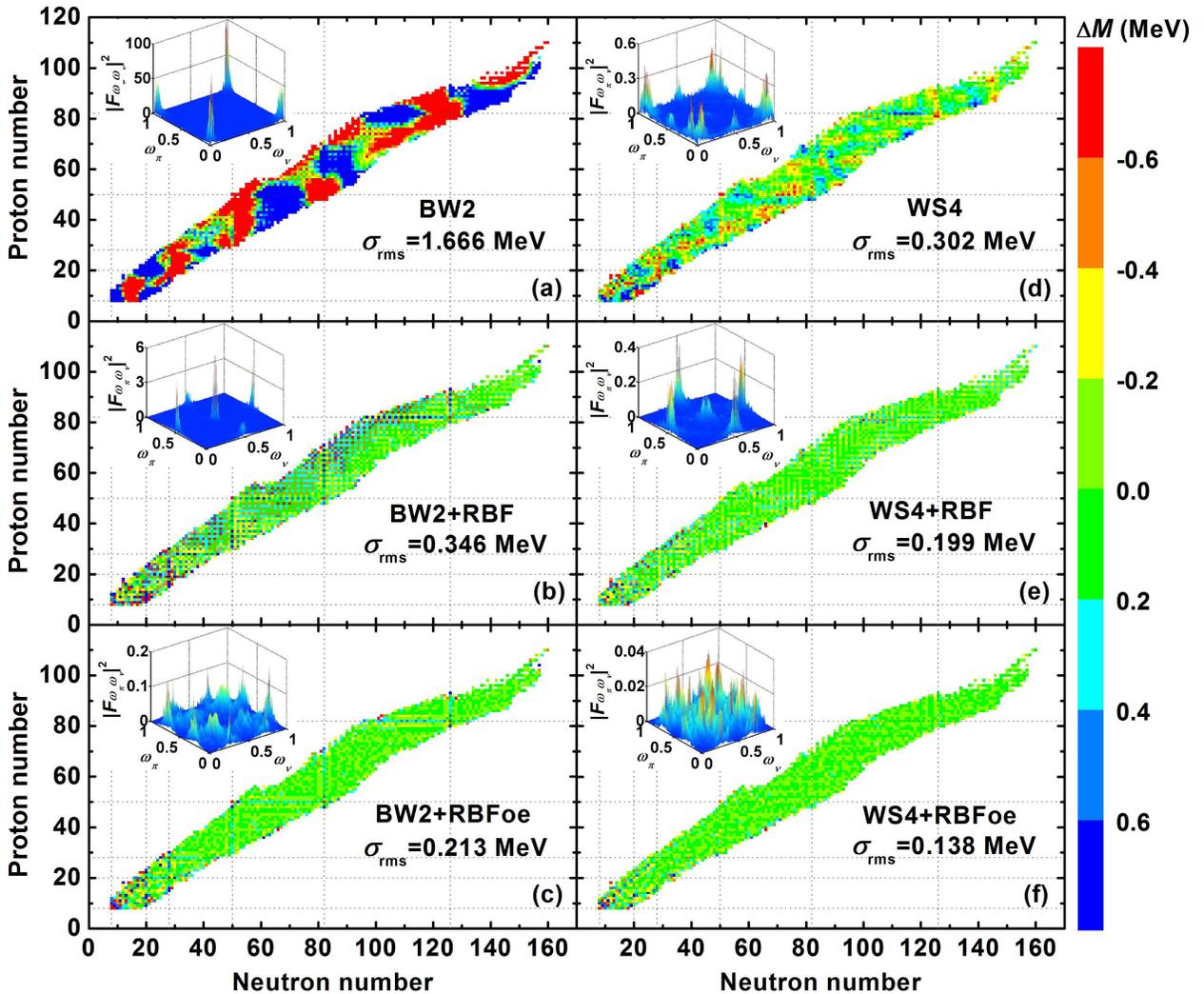}
\caption{(Color online) Impact of the RBF approach on the nuclear mass predictions. Panels (a)-(f) correspond to the mass differences between the experimental data and the predictions of the BW2, BW2+RBF, BW2+RBFoe, WS4, WS4+RBF, and WS4+RBFoe models, respectively. The insets in each panel show the squared amplitudes $|F_{\omega_\pi \omega_\nu}|^2$ of the Fourier transforms of the corresponding mass differences as a function of proton frequency $\omega_\pi$ and neutron frequency $\omega_\nu$.}
\label{fig:fig2}
\end{figure*}

To better understand the dominant peaks located at the low-frequency corners in the squared amplitudes $|F_{\omega_\pi \omega_\nu}|^2$, it is essential to isolate their contributions to the mass deviation from the others. For this purpose, we introduce the RBF approach. It can well describe the smooth surface and has been successfully applied in various engineering fields, even in nuclear mass predictions. Taking the BW2~\cite{Kirson2008NPA} and WS4~\cite{Wang2014PLB} mass models as examples, the mass differences between the experimental data and theoretical predictions are shown as a function of proton and neutron numbers in Figs. 2a and 2d, respectively. The mass deviation surfaces are generally smooth, that is the mass deviations change slowly as a function of neutron and proton numbers in many regions of nuclear chart. Therefore, these mass deviations can be removed in a large extend by the RBF approach, as demonstrated by panels (b) and (e). For simplicity, the BW2 and WS4 models improved by the RBF approach are denoted by BW2+RBF and WS4+RBF hereafter, and their squared amplitudes of the Fourier transforms are given in the insets. Clearly, the low-frequency peaks are well eliminated by using the RBF approach. This indicates that the peaks at low frequencies are strongly correlated with the smoothness of mass deviations over the nuclear chart. The corresponding rms deviations drop from $1.666$ to $0.346$~MeV and from $0.302$ to $0.199$~MeV, for the BW2 and WS4 models, respectively.

After ``corrections" by the RBF approach, the rapid oscillations for the mass deviations appear over the chart as shown in Figs. 2b and 2e. Here, the rapid oscillation means the odd-even staggering of mass deviation, i.e. the mass deviations appear with smaller and larger values alternately. This phenomenon is more visible in the squared amplitudes $|F_{\omega_\pi \omega_\nu}|^2$ that show dominant peaks at the high-frequency regions. From the physics point of view, the odd-even staggering is one of the strongest evidence of the existence of superconductivity in nuclei, also known as nuclear pairing phenomenon~\cite{Bohr1958PR, Broglia2013Book}. Therefore, we can train the RBF separately for different groups of nuclei with different odd-even parities of $(Z, N)$ to remove the oscillations in mass deviations as in Ref.~\cite{Niu2016PRC}. For simplicity, this RBF approach considering the odd-even effects is called the RBFoe approach, and the mass model improved by the RBFoe approach is denoted by model+RBFoe, hereafter. In Figs. 2c and 2f, the mass differences and the corresponding squared amplitudes of the Fourier transforms between the experimental data and predictions of the BW2+RBFoe and WS4+RBFoe models are shown. As expected, the mass deviations related to peaks at high-frequency regions are indeed well eliminated by the RBFoe approach, and the corresponding rms deviations drop from $0.346$ to $0.213$~MeV and from $0.199$ to $0.138$~MeV, respectively. The remaining mass deviations are generally irregular and related to the medium frequencies. The squared amplitude is suppressed by more than two orders of magnitudes for the BW2 and one order of magnitude for the WS4. This demonstrates again the validity and ability of application of the RBF approach to improve mass predictions.

It is important to understand which parts of nuclear mass models play the leading roles in forming the peaks at different frequency regions. This will pave an important guideline for improving the model accuracy and its predictive power. Since the BW2 model is a pure macroscopic mass model, the physics of its different terms can be easily identified. Thus we take the BW2 as an example to illustrate the correlations between the mass deviations related to the peaks at different frequency regions and the underlying physics. The nuclear binding energy in the BW2 model is described as the sum of the $A$-dependent term $E_A$ (including the volume, surface, and curvature terms), the isospin-dependent term $E_I$ (including the symmetry, surface symmetry, and Wigner terms), the Coulomb term $E_C$, the pairing term $E_p$, and the phenomenological shell-correction term $E_{\rm sh}$, i.e.,
\begin{eqnarray}
   E_{\rm tot} = E_A + E_I + E_C + E_p + E_{\rm sh}.
\end{eqnarray}
Their detailed expressions can be found in Ref.~\cite{Kirson2008NPA}.

\begin{figure}[!ht]
\includegraphics[width=6cm]{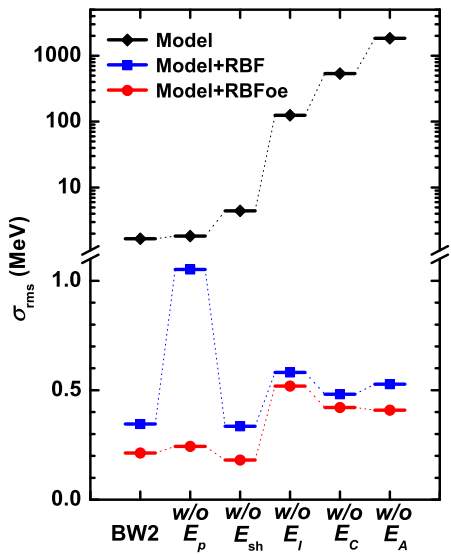}
\caption{(Color online) Effects of the RBF approach on various parts of the BW2 mass model. The short lines with diamonds from the left to right denote the rms deviations of the BW2 model and those respectively removing the pairing term $E_p$, shell-correction term $E_{\rm sh}$, isospin-dependent term $E_I$, Coulomb term $E_C$, and $A$-dependent term $E_A$ from BW2 model. Their corresponding rms deviations improved by the RBF and RBFoe approaches are shown by the short lines with squares and circles, respectively.}
\label{fig:fig3}
\end{figure}

The rms deviations for the BW2 model and the rms deviations calculated by removing respectively various terms $E_A$, $E_I$, $E_C$, $E_p$, and $E_{\rm sh}$ from the BW2 model are shown in Fig. 3. The corresponding results improved by the RBF and RBFoe approaches are also presented. By comparing the rms deviations with and without the RBF improvements, it is clear that the RBF approach significantly reduces the rms deviations, especially for those terms varying smoothly on the nuclear chart, such as the $A$-dependent, Coulomb, and phenomenological shell-correction terms. In fact, better treatment of nuclear mean-field interactions and the global dynamical correlation energies in the mean-field scheme, e.g., rotational and vibrational correction energies~\cite{Lu2015PRC}, is crucial to reduce these deviations, which just correspond to the peaks at low-frequency regions as discussed above. With the RBF improvement, the remaining rms deviation in the case without the pairing term ($E_p=0$), however, is remarkably larger than the other cases. This shows that the pairing effect is difficult to be described by the RBF approach. In such a case, although the RBF approach still works somehow to reduce the rms deviation, it deteriorates the description of single-nucleon separation energy, which indicates the description of pairing interaction is not improved but even becomes worse. This can be understood because the contributions from the pairing interaction lead to the odd-even staggering of nuclear mass, while the RBF approach fails in describing this oscillating behavior. In other words, the high-frequency deviations are due to the improper description of nuclear pairing interaction. With the RBFoe approach, the rms deviations are further reduced and the deviations related to the high frequencies are well eliminated, in particular for the case without the pairing term. Our investigation finds the remaining deviations are mainly related to the peaks at medium frequencies. Another interesting point is that the rms deviation for the case without the isospin-dependent term ($E_I=0$) is relatively larger than other cases after the RBFoe improvements. This implies that a right isospin dependence of nuclear effective interactions is particularly important to reduce the medium-frequency deviations. Comparing with other mass models considered in Fig. 1, the WS4 model has achieved the best accuracy, in which its isospin-dependent components are carefully treated~\cite{Wang2014PLB}. This further supports the importance of isospin-dependent interactions on improving the model accuracy.

\begin{figure}[!ht]
\includegraphics[width=6cm]{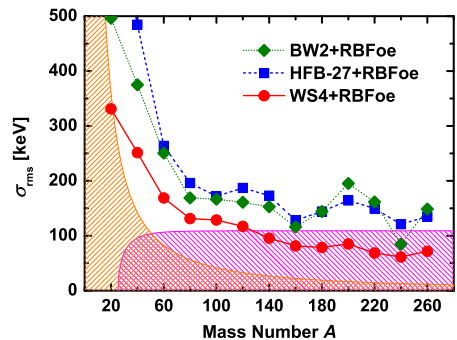}
\caption{(Color online) Mass root-mean-square deviations as a function of mass number $A$ for the BW2+RBFoe, HFB-27+RBFoe, and WS4+RBFoe models. The magenta backslash-hatched and orange slash-hatched areas represent the mass prediction limits induced by the statistical fluctuations of nuclear ground-state energies in Refs.~\cite{Molinari2004PLB} and~\cite{Velazquez2005PLB}, respectively.}
\label{fig:fig4}
\end{figure}

Combining with the RBFoe approach, the accuracy of nuclear mass model is improved significantly. The rms deviation of the WS4+RBFoe model is reduced to $138$~keV, approaching the desired accuracy for the applications in nuclear astrophysics, such as the study of astrophysical nucleosynthesis. Recent studies found that the chaos-related limit for the nuclear mass prediction is also about the same level~\cite{Barea2005PRL}, which implies that the deviations associated to the medium-frequency contributions are of great challenge. To better study this point, we present the rms deviations of the model+RBFoe approaches as a function of mass number $A$ in Fig. 4, while the mass prediction limits~\cite{Molinari2004PLB, Velazquez2005PLB} induced by the statistical fluctuations of nuclear ground-state energies are shown for comparison. The rms deviations of BW2+RBFoe and HFB-27+RBFoe are quite similar, although their original rms deviations are very different. However, their rms deviations are systematically larger than those of WS4+RBFoe. Our investigation finds that the Fourier amplitudes of mass deviations for WS4+RBFoe, HFB-27+RBFoe, and BW2+RBFoe mainly peak at medium frequencies, so the discrepancy between the WS4+RBFoe and HFB-27+RBFoe or BW2+RBFoe models indicates the deviations causing the peaks at medium frequencies can be further decreased by better treating the nuclear effective interactions as in WS4 model. Therefore, better isospin dependence of effective interactions in WS4 model is helpful to improve the nuclear mass predictions. For the WS4+RBFoe model, its rms deviation crosses the $100$-keV accuracy when $A \gtrsim 120$, which is lower than the accuracy limit estimated in Ref.~\cite{Molinari2004PLB}, while larger than the one estimated in Ref.~\cite{Velazquez2005PLB}. However, the rms deviation is larger than the accuracy limits in both Refs.~\cite{Molinari2004PLB} and~\cite{Velazquez2005PLB} when $A \lesssim 120$. Since the mean-field approximation is relatively poor for nuclei with small $A$, those neglected many-body correlations in the mean-field calculations are helpful to further decrease the mass deviations peaked at medium frequencies, which probably can be treated effectively by including more reliable isospin dependence of nuclear effective interactions.

\begin{figure}[!ht]
\includegraphics[width=6cm]{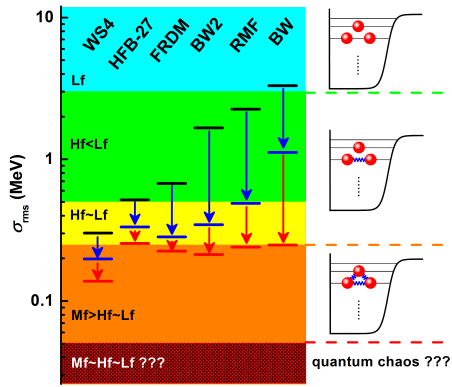}
\caption{(Color online) Schematic illustration of the relations between the accuracy desired for nuclear mass model and the required nuclear effective interactions to be properly considered. In each column the lines from top to bottom denote the mass rms deviations for the model, model+RBF, and model+RBFoe approaches. Models from left to right denote the WS4~\cite{Wang2014PLB}, HFB-27~\cite{Goriely2013PRC}, FRDM~\cite{Moller1995ADNDT}, BW2~\cite{Kirson2008NPA}, RMF~\cite{Geng2005PTP}, and BW~\cite{Bethe1936RMP} models, respectively.}
\label{fig:fig5}
\end{figure}

So far we conclude that the nuclear mean-field interactions and the dynamical corrections are crucial for mass calculations. If they are not properly described theoretically, it will cause mass deviations demonstrated as the sharp peaks at low frequencies (Lf) in the conjugate space of frequency, while the peaks emerged at high frequencies (Hf) are associated with the mass deviations induced by improper treatment of pairing interactions. Moreover, the neglected many-body correlations in the mean-field calculations are helpful to decrease the mass deviations corresponding to the peaks at medium frequencies (Mf), which may be effectively treated by considering more reliable isospin dependence of nuclear effective interactions. Here we present a strategy to improve the nuclear mass model, namely, by using the RBF and RBFoe approaches to minimize the possible ``noise" at high and low frequencies. Taking the WS4, HFB-27, FRDM, BW2, RMF, and BW models as examples, the rms deviations and those improved by the RBF and RBFoe approaches are given in Fig. 5. Note that there is no pairing term in the BW model, its $\sigma_{\textrm{rms}}\gtrsim$ is still about $3$~MeV. By including the pairing interaction, the rms deviations are reduced, while the $\sigma_{\textrm{rms}}$ of RMF and BW2 models are still larger than $1$~MeV, and their squared amplitudes of the Fourier transforms concentrate at the low-frequency regions. With the RBF approach, the deviations in the low-frequency regions are well eliminated, and the $\sigma_{\textrm{rms}}$ of RMF model is reduced to about $500$~keV. That is, for developing a mass model with $500$~keV $\lesssim \sigma_{\textrm{rms}} \lesssim$ $3$~MeV, the pairing interaction is necessary, while mass deviations still induce the peaks at low frequencies, which can be reduced by better treating nuclear mean-field interactions and the dynamical corrections. With the RBFoe approach, the deviations in the high-frequency regions are further reduced. The rms deviations for different models are all reduced to around $250$ keV except the WS4 model, whose rms deviation is reduced to $138$ keV. The isospin-dependent components in the WS4 model is carefully treated and would help to decrease mass deviations at the medium frequencies. Thus, for developing a mass model with $250$~keV $\lesssim \sigma_{\textrm{rms}} \lesssim$ $500$~keV, the nuclear pairing interaction, mean-field interactions and the dynamical corrections are almost equally important. However, for achieving an accuracy better than about $250$~keV, the many-body correlations become crucial to decrease the mass deviations peaked at the medium frequencies, which probably can be treated by including isospin dependence of nuclear effective interactions. Finally, when the rms deviations reduce to the chaos-related accuracy limit, the corresponding mass deviations may behave like white noise, whose frequency distribution is totally irregular and hence no particular frequency is dominated. This would be a great challenge to accurate quantum many-body calculations. Other nuclear mass models, such as the DZ10~\cite{Duflo1995PRC}, DZ31~\cite{Zuker2008RMFS}, ETFSI-Q~\cite{Pearson1996PLB}, and KTUY~\cite{Koura2005PTP} mass models, are also employed to verify our conclusions.

\section{Summary and perspectives}\label{Sec:4}
In this work, we offer a new insight to improve the accuracy of nuclear mass model by combining the sophisticated techniques in engineering --- the Fourier spectral analysis and the radial basis function approach. A guiding way for improving the accuracy of nuclear mass predictions from several million electron volts to the chaos-related accuracy limit is specified: The mean-field interactions and the dynamical corrections play a leading role for achieving an accuracy of $500$~keV $\lesssim \sigma_{\textrm{rms}} \lesssim$ $3$~MeV; The pairing interaction is necessary for achieving an accuracy better than about $3$~MeV, and becomes crucial for an accuracy better than about $500$~keV; In order to have an accuracy better than about $250$~keV, the neglected many-body correlations in the mean-field calculations are essential, which probably can be treated by including isospin dependence of nuclear effective interactions. This new insight paves a general and systematic way for developing a nuclear mass model towards a hundred kilo-electron-volt accuracy, which will be the key for many unsolved fundamental questions, such as the origin of heavy elements in the universe.

\section*{Conflicts of interest}
The authors declare that they have no conflict of interest.

\section*{Acknowledgements}
This work was supported by the Major State 973 Program of China (Grant No. 2013CB834400), the National Natural Science Foundation of China (Grant No. 11205004, No. 11305161, No. 11335002, No. 11475014, No. 11575002, and No. 11411130147), the Natural Science Foundation of Anhui Province under Grant No.~1708085QA10, and the RIKEN iTHES project and iTHEMS program.




\end{document}